\documentclass[apjl]{emulateapj}
\shorttitle{Abundance of Void Groups}
\shortauthors{Song \& Lee}
\begin{document}
\title{The Mass Function of Void Groups as a Probe of the Primordial 
non-Gaussianity}
\author{Hyunmi Song and Jounghun Lee}
\affil{Department of Physics and Astronomy, FPRD, Seoul National University, 
Seoul 151-747, Korea} 
\email{yesuane@gmail.com}
\email{jounghun@astro.snu.ac.kr}
\begin{abstract}
The primordial non-Gaussianity signal, if measured accurately, will allow us
to distinguish between different candidate models for the cosmic inflation.
Since the galaxy groups located in void regions are rare events, their 
abundance may be a sensitive probe of the primordial non-Gaussianity.
We construct an analytic model for the mass function of void groups in the 
framework of the extended Press-Schechter theory with non-Gaussian initial 
conditions and investigate how it depends on the primordial non-Gaussianity 
parameter. A feasibility study is conducted by fitting the analytic mass 
function of void groups to the observational results from the galaxy 
group catalog of the Sloan Digital Sky Survey Data Release 4 with adjusting 
the primordial non-Gaussianity parameter.
\end{abstract}
\keywords{cosmology:theory --- large-scale structure of universe}
%%%%%%%%%%%%%%%%%%%%%%%%%%%%%%%%%%%%%%%%%%%%%%%%%%%%%%%%%%%%%%%%%%%%%%%%%%%

\section{INTRODUCTION}

In precision cosmology, a fundamental task is to distinguish between various 
candidate models for the cosmic inflation. One way to perform this task is to 
measure the degree of non-Gaussianity in the primordial density field. 
Although the primordial density field is regarded as nearly Gaussian in all 
inflationary scenarios \citep{GP82}, the degree of its non-Gaussianity 
differs between the models. For instance, in the single-field slow-roll 
inflationary model, the deviation from Gaussianity is small enough 
to be unobservable \citep{ver-etal01}. Meanwhile in some multi-field 
inflationary models the non-Gaussianity can be generated to a detectable 
level \citep[][and references therein]{BE07}. For a thorough 
review on the predictions of various inflationary models for the 
primordial non-Gaussianity, see \citet{bar-etal04}

The cosmic microwave background radiation (CMB) is a useful probe of the 
primordial non-Gaussianity as it reflects the linear density perturbations. 
Yet, if the primordial non-Gaussianity is scale-dependent, the constraint 
from the CMB analysis is rather restricted to the very large-scale 
\citep{ver-etal01,KS01}.  
The large-scale structure is a powerful alternative probe of the primordial 
non-Gaussianity on the sub-CMB scale. It has been well known that the 
abundance of high-$z$ clusters are rare enough to constrain the primordial 
non-Gaussianity \citep{LM88,chi-etal98,mat-etal00,WK03,car-etal08,lov-etal08,
gro-etal09}. This probe, however, is likely to suffer from large systematics 
involved in the inaccurate measurement of the masses of high-$z$ clusters. 
The abundance of cosmic voids is another probe of the primordial 
non-Gaussianity based on the large-scale structure \citep{kam-etal09}. 
One difficulty in using this probe lies in the fact that there is no unique 
way to define voids \citep{car-etal08}. Recently, it has been claimed that 
using the clustering properties of highly biased large-scale structures 
the primordial non-Gaussianity parameter can be measured with accuracy 
as high as the one obtained from the CMB analysis 
\citep{car-etal08,slo-etal08,jeo-kom09}. 

Here, we propose the mass function of present galaxy groups embedded in 
void regions as a new probe of the primordial non-Gaussianity. It is not only 
the high-$z$ clusters but also the low-$z$ void groups that are so rare that 
their abundance may depend sensitively on the initial conditions. 
Furthermore, the mass estimation of low-$z$ galaxy groups should be much 
more reliable than that of high-$z$ clusters \citep{yan-etal07}. 
Throughout this Letter, we assume a WMAP 5 cosmology \citep{wmap5}.

\section{AN ANALYTIC MODEL}

The Press-Schechter theory \citep[][PS hereafter]{PS74} provides an analytic 
framework within which the number density of bound objects as a function of 
mass, $dN_{\rm PS}/dM$, can be obtained:
\begin{equation}
\label{eqn:ps}
\frac{dN}{dM} = A\frac{\bar{\rho}}{M}\frac{d}{dM}\left\vert
\int_{\delta_c(z)}^{\infty}p(\delta_{M})d\delta\right\vert,
\end{equation} 
where $\bar{\rho}$ is the mean background density, $\delta_{c}(z)$ is the 
critical density contrast for gravitational collapse at redshift $z$, 
$p(\delta_{M})$ is the probability density distribution of the density field 
$\delta_{M}$ smoothed on the mass scale of $M$, and $A$ is the normalization 
constant. Basically, equation (\ref{eqn:ps}) states that the number density 
of bound objects can be inferred from the differential volume fraction 
occupied by those regions in the linear density field whose average density 
contrast $\delta_{M}$ reaches a certain threshold, $\delta_{c}$.
In the original PS theory, the initial density field is assumed to be 
Gaussian as $p(\delta_{M})=\exp\left[-\delta^{2}_{M}/(2\sigma^{2}_{M})\right]
/(\sqrt{2\pi}\sigma_{M})$. 

Here $\sigma_{M}$ is the rms density fluctuation 
smoothed on the mass scale $M$ and $\delta_{c}\equiv\delta_{c0}/D_{+}(z)$ 
where $\delta_{c0}$ is the critical density contrast at $z=0$ and $D_{+}(z)$ 
is the linear growth factor. For a WMAP 5 cosmology, we find 
$\delta_{c0}\approx 1.62$. Note that the mass function of bound objects 
depends on the initial conditions through its dependence on $\sigma_{M}$ 
which is a function of the density parameter, $\Omega_{m}$ and the 
amplitude of the linear power spectrum, $\sigma_{8}$. 

Now, let us consider the case of non-Gaussian initial conditions that is 
often characterized by the primordial non-Gaussianity parameter $f_{\rm NL}$ 
as  $\psi({\bf x})=\phi({\bf x}) + f_{NL}\left(\phi^{2}({\bf x})-
\langle\phi^{2}({\bf x})\rangle\right)$ where $\phi$ is a Gaussian random 
field and $\psi({\bf x})$ is the linearly extrapolated gravitational 
potential at $z=0$ \citep[e.g.,][]{lov-etal08,gro-etal09}. 
The functional form of $p(\delta_{M})$ for the 
non-Gaussian case will be in general different from the Gaussian case. 
However, provided that the degree of the non-Gaussianity is very small and 
scale independent, $p(\delta_{M})$ has the same functional form as the 
Gaussian case at first order \citep{LM88,mat-etal00,ver-etal01}. 
The only difference is the value of the critical density contrast, 
$\delta_{c*}$, which is related to that of the Gaussian case, $\delta_{c}$ as 
\begin{equation}
\label{eqn:nldc}
\delta_{c*}(z)=\delta_{c}(z)\left[1-\frac{S_3}{3}\delta_{c}(z)\right]^{1/2}.
\end{equation}
Here $S_3$ is a skewness parameter, related to the primordial non-Gaussianity 
parameter $f_{\rm NL}$ as $S_{3}=\frac{3}{5}f_{\rm NL}\mu_{3}^{(1)}/
(\mu_{2}^{(1)})^2$ where $\mu_{2}^{(1)}$ and $\mu_{3}^{(1)}$ denote the 
variance and skewness of the smoothed non-Gaussian density field at first 
order, respectively  \citep[see eqs.43-45 in][]{mat-etal00}. 
Therefore, the PS mass function with non-Gaussian initial conditions is a 
function of $M$ and $f_{\rm NL}$: $dN_{\rm PS}(f_{\rm NL},M)/dM$. The 
case of $f_{\rm NL}=0$ corresponds to the original one with Gaussian initial 
conditions.

On the group scale, $dN_{\rm PS}(f_{\rm NL},M)/dM$ does not change sensitively 
with the initial conditions since the galaxy groups are not rare events. 
However, those groups located in voids should be so rare that their 
abundance may depend sensitively on the initial conditions. Before deriving 
the abundance of void groups, we clarify the meaning of a {\it void region}. 
Following \citet{hah-etal07}, we define a void region on mass scale $M$ as a 
region where the three eigenvalues of the tidal tensor at a given region on 
mass scale $M$ are less than zero. Then, we replace $p(\delta_{M})$ in 
eq.~(\ref{eqn:ps}) by $p(\delta_{M}|\lambda^{\prime}_{1M^{\prime}}<0)$ which 
represents the conditional probability density distribution that the density 
contrast has a certain value on the mass scale $M$ provided that the largest 
eigenvalue of the tidal tensor is negative on some larger mass scale 
$M^{\prime}>M$. From here on, $\delta$ denotes the density contrast on the 
mass scale $M$, while $\lambda^{\prime}_{1},\ \lambda^{\prime}_{2},\ 
\lambda^{\prime}_{3}$ are the three eigenvalues of the tidal field 
on some larger mass scale $M^{\prime}$.

The joint probability distribution 
$p(\delta,\lambda^{\prime}_{1},\lambda^{\prime}_{2},\lambda^{\prime}_{3})$ 
for the case of Gaussian initial conditions has been already derived by 
\cite{lee06} as  
\begin{eqnarray}
\label{eqn:ess}
p(\delta,\lambda^{\prime}_{1},\lambda^{\prime}_{2},
\lambda^{\prime}_{3}) = 
\frac{1}{\sqrt{2\pi}\sigma_{\Delta}}\frac{3375}{8\sqrt{5}\pi\sigma^{\prime6}}
\exp\left[-\frac{(\delta-I^{\prime}_{1})^{2}}
{2\sigma^{2}_{\Delta}}\right]\times \nonumber \\
\exp\left(-\frac{3I^{\prime 2}_{1}}{\sigma^{\prime 2}} + 
\frac{15I^{\prime}_{2}}{2\sigma^{\prime 2}}\right)
(\lambda^{\prime}_{1}-\lambda^{\prime}_{2})
(\lambda^{\prime}_{2}-\lambda^{\prime}_{3})
(\lambda^{\prime}_{1}-\lambda^{\prime}_{3}), 
\end{eqnarray}
with $\sigma^{2}_{\Delta} \equiv \sigma^{2} - \sigma^{\prime 2}$, 
$I^{\prime}_{1}=\lambda^{\prime}_{1} + \lambda^{\prime}_{2} + 
\lambda^{\prime}_{3}$, and $I^{\prime}_{1}=\lambda^{\prime}_{1}
\lambda^{\prime}_{2}+\lambda^{\prime}_{2}\lambda^{\prime}_{3}+ 
\lambda^{\prime}_{1}\lambda^{\prime}_{3}$. Here $\sigma$ and 
$\sigma^{\prime}$ represents the rms density fluctuations on the mass 
scale $M$ and $M^{\prime}$, respectively. 
The conditional probability density, $p(\delta| \lambda^{\prime}_{1}<0)$, is 
now written as 
\begin{equation}
\label{eqn:void}
p(\delta| \lambda^{\prime}_{1}<0) = 
\frac{\int^{0}_{-\infty}d\lambda_{1}
\int^{\lambda^{\prime}_{1}}_{-\infty}d\lambda_{2}
\int^{\lambda^{\prime}_{2}}_{-\infty}d\lambda_{3}
p(\delta,\lambda^{\prime}_{1},\lambda^{\prime}_{2},\lambda^{\prime}_{3})}
{\int^{0}_{-\infty}d\lambda_{1}
\int^{\lambda^{\prime}_{1}}_{-\infty}d\lambda_{2}
\int^{\lambda^{\prime}_{2}}_{-\infty}d\lambda_{3}
p(\lambda^{\prime}_{1},\lambda^{\prime}_{2},\lambda^{\prime}_{3})}.
\end{equation} 
Putting $p(\delta|\lambda^{\prime}_{1}<0)$ in eq.~(\ref{eqn:ps}), we evaluate 
the mass function of void groups with Gaussian initial conditions, 
$dN_{\rm V}^{\rm G}/dM$.
\begin{figure} 
\plotone{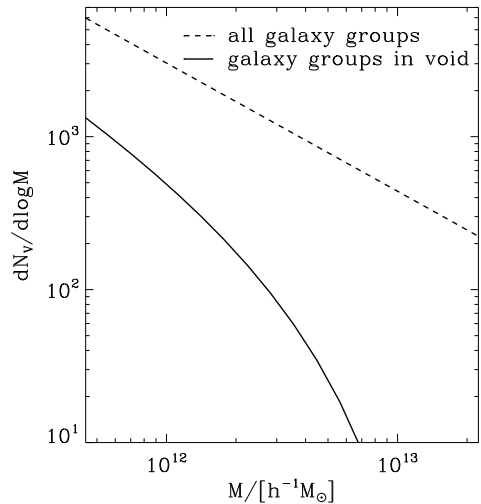}
\caption{Mass function of void groups (solid line) and of all 
groups (dashed line) in the extended Press-Schechter formalism.}
\label{fig:analytic}
\end{figure}

Using the methodology suggested by \citet{lov-etal08}, we model the mass 
function of void groups with non-Gaussian initial conditions, 
$dN^{\rm NG}_{\rm V}/dM$, as
\begin{equation}
\label{eqn:ratio}
\frac{dN^{\rm NG}_{\rm V}}{dM} =
\frac{dN^{\rm G}_{\rm V}}{dM}\frac{dN_{\rm PS}(f_{\rm NL},M)/dM}
{dN_{\rm PS}(f_{\rm NL}=0,M)/dM}, 
\end{equation}
where $dN_{\rm PS}(f_{\rm NL}=0,M)/dM$ and $dN_{\rm PS}(f_{\rm NL},M)/dM$ 
represents the PS mass function  with Gaussian and non-Gaussian initial 
conditions, respectively (see eq.[4.20] in Loverde et al. 2009). 
Figure \ref{fig:analytic} plots the mass function of void groups (solid line), 
which normalized as $\int dN_{V}/d\log M_{i}=N_{vg}$ where $N_{vg}$ is the 
total number of void groups found in the observational data (see \S 3.)
The mass function of all groups (dashed line) are also plotted for comparison. 
As can be seen, the mass function of void groups decreases very rapidly with 
mass, which indicates that it must depend sensitively on the initial 
conditions. 

It has to be mentioned here that equation (\ref{eqn:ratio}) has yet to be 
validated against numerical results. The methodology of \citet{lov-etal08} 
that equation (\ref{eqn:ratio}) is based on has recently been tested against 
N-body simulations and found to be valid in the high-mass sections 
\citep{gro-etal09}. Yet, to fully justify the use of equation 
(\ref{eqn:ratio}) for the evaluation of the abundance of void groups with 
non-Gaussian initial conditions, it will be required to test equation 
(\ref{eqn:ratio}) numerically on the group-mass scale. 
Furthermore, for the case of scale-dependent non-Gaussianity the mass 
function of void groups would have much more complicated even at first order 
\citep{lov-etal08}. The focus of this work, however, is on the proof of a 
concept that the abundance of present galaxy groups in voids can be used 
as a probe of the primordial non-Gaussianity. Henceforth, here we use 
equation (\ref{eqn:ratio}) as an ansatz and consider only the 
scale-independent case for simplicity.
\begin{figure} 
\plotone{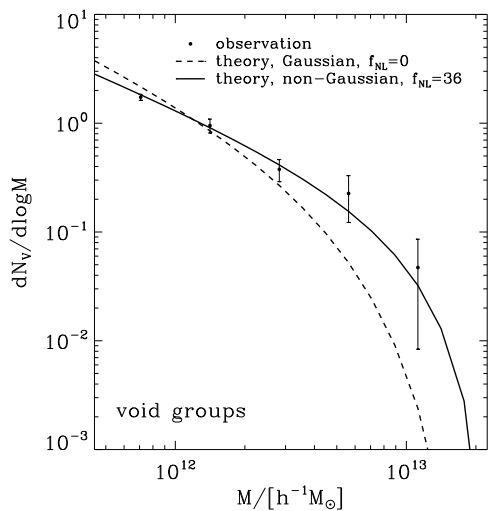}
\caption{Mass function of void groups with the best-fit value of 
$f_{\rm NL}$ (solid line) determined by fitting to the observational data 
from the SDSS DR4 (dots). The errors include both the Poisson noise and the 
cosmic variance. The mass function of void groups for the Gaussian 
case ($f_{\rm NL}=0$) is also plotted for comparison (dashed).}
\label{fig:void}
\end{figure}

\section{A FEASIBILITY STUDY}

We conduct a feasibility study by comparing the analytic mass function of 
void groups obtained in \S 2 with the observational result from the 
galaxy group catalog of the Sloan Digital Sky Survey Data Release 4 
provided by \citep{yan-etal07} who measured the masses of the SDSS galaxy 
groups were estimated from the characteristic luminosity (or stellar masses) 
using the WMAP 3 cosmology \citep{wmap3}.
To identify void groups from the SDSS group catalog, we use the real-space 
tidal field reconstructed by \citet{LE07} on $64^{3}$ pixels in a box of 
linear length $400h^{-1}$Mpc from the Two Mass Redshift survey (2MRS) 
\citep{huc-etal05}. We smooth the 2MRS tidal field with a Gaussian filter of 
scale radius $8h^{-1}$Mpc on which scale the density field is still in the 
quasi-linear regime, calculate the three eigenvalues of the smoothed tidal 
field at each pixel and mark as voids those pixels in which all three 
eigenvalues are less than zero. 
A total of $550$ galaxy groups at redshifts $0.01\le z<0.04$ in mass range of 
$11.7\le\log M/(h^{-1}M_{\odot})\le 13.4$ are found in the void pixels. 
Binning their mass range in the logarithmic scale, we measure their abundance,
$dN_{V}/d\log M$, which is renormalized to be $\int dN_{V}=1$. 
To account for the cosmic variance as well as the Poisson noise in the 
measurement of $dN_{V}/d\log M$, we divide the selected void groups into 
$6$ subsamples and measure  $dN_{V}/d\log M$ for each subsample separately. 
We calculate the jackknife errors as the standard deviation in the measurement 
of the mean $dN_{V}/d\log M$ averaged over the $6$ subsamples at each mass bin.

We fit the observational result to the analytic model by adjusting the 
value of $f_{\rm NL}$. To account for the correlations between the mass bins, 
we employ the {\it generalized} $\chi^{2}$-statistics to determine the 
best-fit value of $f_{\rm NL}$: $\chi^{2} = [n_{i}-n(\log M_{i};f_{\rm NL})]
C^{-1}_{ij}[n_{i}-n(\log M_{i};f_{\rm NL})]$ where $n_{i}\equiv dN/d\log M_{i}$
and $n(\log M_{i};f_{\rm NL})$ represents the observational and analytical 
results evaluated at the $i$-th logarithmic mass bin, $\log M_{i}$, 
respectively. And $(C_{ij})$ is the covariance matrix defined as 
$C_{ij}=\langle(n_{i}-n_{0i})(n_{j}-n_{0j})\rangle$ where $n_{0i}$ 
represents the mean of $n_{i}$ averaged over all samples. 
Finally, the uncertainty in the measurement of $f_{\rm NL}$ is calculated as 
the curvature of the $\chi^{2}$ function at the minimum. Through this fitting 
procedure, we find $f_{\rm NL}=36\pm 1$. Figure \ref{fig:void} plots the 
observational results (dots) and the analytic model with the best-fit value 
of $f_{\rm NL}=36$ (solid line). The median redshift of the SDSS void 
groups, $z=0.03$, is used for the value of $z$ in the analytic model.  
For comparison, the analytic model with $f_{\rm NL}=0$ is also plotted 
(dotted line). As can be seen, the observational results agree better with 
the analytic model with non-Gaussian initial conditions. 

\begin{figure} 
\plotone{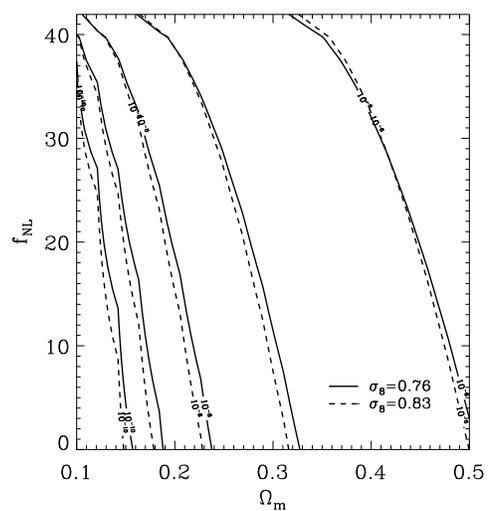}
\caption{Degeneracy curves in the $f_{\rm NL}$-$\Omega_{\rm m}$ plane with 
the value of $\sigma_{8}$ set at $0.76$ (solid)and at $0.83$ (dashed).}
\label{fig:con}
\end{figure}

It is, however, expected that there is a degeneracy between $f_{\rm NL}$ and 
the other key cosmological parameters on which the mass function of void 
groups depend. Here, we investigate the degeneracy between $\Omega_{m}$ 
and $f_{\rm NL}$, setting $\sigma_{8}$ at the value determined by 
WMAP 5 cosmology. Varying the values of $\Omega_{m}$ and $f_{\rm NL}$, 
we recalculate $dN_{V}/d\log M$ at a typical group mass scale of 
$M=10^{13}h^{-1}M_{\odot}$.  Figure \ref{fig:con} plots a family of the 
degeneracy curves in the $\Omega_m$-$f_{\rm NL}$ plane with the value 
of $\sigma_{8}$ set at $0.76$ (solid) and $0.83$ (dashed). As can be seen, 
a strong degeneracy exists between the two parameters. For a fixed value 
of $dN_{V}/d\log M$, the value of $f_{\rm NL}$ increases as the value of 
$\Omega_{m}$ decreases. The comparison between the solid and the dashed lines 
indicate that the degree of the degeneracy between $\Omega_{m}$ and 
$f_{\rm NL}$ increases as the value of $\sigma_{8}$ increases. 
It also indicates that if $\Omega_{m}$ is fixed, the value of 
$f_{\rm NL}$ increases as the value of $\sigma_{8}$ decreases.
To break this parameter degeneracy, it will be necessary to combine 
our analysis with other analyses.

The preliminary results of this feasibility study, however, are subject to 
several caveats. First, the analytic model assumes the scale-independent 
Gaussianity. To be more realistic, it is necessary to account for the 
scale-dependent non-Gaussianity. The second caveat lies in the limitation of 
the Press-Schechter formalism. As shown by several authors 
\citep{LS98,ST99,jen-etal01}, the real gravitational process deviates from 
the spherical dynamics on which the PS mass function is based. It has to 
be tested how significantly the deviation of gravitational collapse process 
from the spherical dynamics affects the abundance of void groups.  
The third caveat comes from the fact that the validity of equation 
(\ref{eqn:ratio}) has yet to be confirmed.  Although \citet{gro-etal09} have 
shown that this methodology suggested by \citet{lov-etal08} to model 
departures from non-Gaussianity leads to an excellent approximation on the 
cluster scale, it has to be confirmed by N-body simulations whether the 
same methodology can be used to count the number of void groups with 
non-Gaussian initial conditions. Fourth, the different mass-to-light ratios 
of the void galaxies from that of the wall galaxies has to be taken into 
account. According to \citet{roj-etal05}, the specific star formation rate 
in void galaxies is higher than that in wall galaxies, suggesting that the 
mass of the void groups in the SDSS Galaxy group catalog are likely to be 
overestimated. Fifth, \citet{yan-etal07} measured the masses of SDSS galaxy 
groups assuming the old WMAP 3 cosmology \citep{wmap3}. It will be necessary 
to use the values of the most updated WMAP 5 parameters for the more accurate 
calculation of the masses of void groups.

As a final conclusion, we have proved that the abundance of void groups 
can in principle be a useful probe of the primordial non-Gaussianity 
parameter. For a robust probe, however, it will be required to refine 
further the analytic model of the abundance of void groups and 
to improve the mass estimation of galaxy groups in void regions, 
which is the direction of our future work.

\acknowledgments

We thank an anonymous referee for many helpful suggestions.
This work is financially supported by the Korea Science and Engineering 
Foundation (KOSEF) grant funded by the Korean Government 
(MOST, NO. R01-2007-000-10246-0).

Funding for  the SDSS and SDSS-II  has been provided by  the Alfred P.
Sloan Foundation, the Participating Institutions, the National Science
Foundation, the  U.S.  Department of Energy,  the National Aeronautics
and Space Administration, the  Japanese Monbukagakusho, the Max Planck
Society, and  the Higher Education  Funding Council for  England.  The
SDSS Web  Site is  http://www.sdss.org/.  The SDSS  is managed  by the
Astrophysical    Research    Consortium    for    the    Participating
Institutions. The  Participating Institutions are  the American Museum
of  Natural History,  Astrophysical Institute  Potsdam,  University of
Basel,   Cambridge  University,   Case  Western   Reserve  University,
University of Chicago, Drexel  University, Fermilab, the Institute for
Advanced   Study,  the  Japan   Participation  Group,   Johns  Hopkins
University, the  Joint Institute  for Nuclear Astrophysics,  the Kavli
Institute  for   Particle  Astrophysics  and   Cosmology,  the  Korean
Scientist Group, the Chinese  Academy of Sciences (LAMOST), Los Alamos
National  Laboratory, the  Max-Planck-Institute for  Astronomy (MPIA),
the  Max-Planck-Institute  for Astrophysics  (MPA),  New Mexico  State
University,   Ohio  State   University,   University  of   Pittsburgh,
University  of  Portsmouth, Princeton  University,  the United  States
Naval Observatory, and the University of Washington.


\begin{thebibliography}{}
\bibitem[Battefeld \& Easther(2007)]{BE07} 
Battefeld, T., \& Easther, R.\ 2007, Journal of Cosmology and Astro-Particle 
Physics, 3, 20 
\bibitem[Bartolo et al.(2004)]{bar-etal04} 
Bartolo, N., Komatsu, E., Matarrese, S., \& Riotto, A.\ 2004, 
\physrep, 402, 103 
\bibitem[Chiu et al.(2998)]{chi-etal98} 
Chiu, W.~A., Ostriker, J. P., \& Strauss, M.~A.\ 1998, \apj, 494, 479 
\bibitem[Carbone et al.(2008)]{car-etal08} 
Carbone, C., Verde, L., \& Matarrese, S.\ 2008, \apjl, 684, L1 
\bibitem[Colberg et al.(2008)]{col-etal08} 
Colberg, J.~M., et al.\ 2008, \mnras, 387, 933 
\bibitem[Dunkley et al.(2009)]{wmap5} 
Dunkley, J., et al.\ 2009, \apjs, 180, 306 
\bibitem[Eke et al.(1996)]{eke-etal96} 
Eke, V.~R., Cole, S., \& Frenk, C.~S.\ 1996, \mnras, 282, 263 
\bibitem[Erdo{\u g}du et al.(2006)]{erd-etal06} Erdo{\u g}du, P., et al.
\ 2006, \mnras, 373, 45 
\bibitem[Grossi et al.(2009)]{gro-etal09} 
Grossi, M., Verde, L., Carbone, C., Dolag, K., Branchini, E., Iannuzzi, F., 
Matarrese, S., \& Moscardini, L.\ 2009, arXiv:0902.2013 
\bibitem[Guth \& Pi(1982)]{GP82} 
Guth, A.~H., \& Pi, S.-Y.\ 1982, \prl, 49, 1110 
\bibitem[Hahn et al.(2007)]{hah-etal07} 
Hahn, O., Carollo, C.~M., Porciani, C., \& Dekel, A.\ 2007, \mnras, 381, 41 
\bibitem[Huchra et al.(2005)]{huc-etal05} Huchra, J., et al.\ 
2005, Maps of the Cosmos, 216, 170 
\bibitem[Jenkins et al.(2001)]{jen-etal01}
Jenkins, A., Frenk, C.~S., White, S.~D.~M., Colberg, J.~M., Cole, S., 
Evrard, A.~E., Couchman, H.~M.~P., \& Yoshida, N.\ 2001, \mnras, 321, 372
\bibitem[Jeong \& Komatsu(2009)]{jeo-kom09} 
Jeong, D., \& Komatsu, E.\ 2009, arXiv:0904.0497 
\bibitem[Kamionkowski et al.(2009)]{kam-etal09} 
Kamionkowski, M., Verde, L., \& Jimenez, R.\ 2009, Journal of Cosmology 
and Astro-Particle Physics, 1, 10 
\bibitem[Komatsu \& Spergel(2001)]{KS01} 
Komatsu, E., \& Spergel, D.~N.\ 2001, \prd, 63, 063002 
\bibitem[Lee(2006)]{lee06}
Lee, J. 2006, preprint (arXiv:0605697v1)
\bibitem[Lee \& Erdogdu(2007)]{LE07} 
Lee, J., \& Erdogdu, P.\ 2007, \apj, 671, 1248 
\bibitem[Lee \& Shandarin(1998)]{LS98}
Lee, J. \& Shandarin, S.~F.\ 1998, ApJ, 500, 14
\bibitem[LoVerde et al.(2008)]{lov-etal08} 
Lo Verde, M., Miller, A., Shandera, S., \& Verde, L.\ 2008, 
Journal of Cosmology and Astro-Particle Physics, 4, 14 
\bibitem[Lucchin \& Matarrese(1988)]{LM88} 
Lucchin, F., \& Matarrese, S.\ 1988, \apj, 330, 535 
\bibitem[Matarrese et al.(2000)]{mat-etal00} 
Matarrese, S., Verde, L., \& Jimenez, R.\ 2000, \apj, 541, 10 
\bibitem[Press \& Schechter(1974)]{PS74} 
Press, W.~H., \& Schechter, P.\ 1974, \apj, 187, 425 
\bibitem[Rojas et al.(2005)]{roj-etal05}
Rojas, R.~R., Vogeley, M.~S., Hoyle, F., \& Brinkmann, J.\ 2005, 624, 571
\bibitem[Sheth \& Tormen(1999)]{ST99}
Sheth, R.~K., \& Tormen, G.\ 1999, \mnras, 308, 119 
\bibitem[Slosar et al.(2008)]{slo-etal08} 
Slosar, A., Hirata, C., Seljak, U., Ho, S., \& Padmanabhan, N.\ 2008, 
Journal of Cosmology and Astro-Particle Physics, 8, 31 
\bibitem[Spergel et al.(2007)]{wmap3}
Spergel, D.~N., et al.\ 2007, \apjs, 170, 377 
\bibitem[Verde et al.(2000)]{ver-etal00} 
Verde, L., Wang, L., Heavens, A.~F., \& Kamionkowski, M.\ 2000, \mnras, 
313, 141 
\bibitem[Verde et al.(2001)]{ver-etal01} 
Verde, L., Jimenez, R., Kamionkowski, M., \& Matarrese, S.\ 2001, 
\mnras, 325, 412 
\bibitem[Weinberg \& Kamionkowski(2003)]{WK03} 
Weinberg, N.~N., \& Kamionkowski, M.\ 2003, \mnras, 341, 251 
\bibitem[Yang et al.(2007)]{yan-etal07} 
Yang, X., Mo, H.~J., van den Bosch, F.~C., Pasquali, A., Li, C., 
\& Barden, M.\ 2007, \apj, 671, 153 

\end{thebibliography}
\end{document}